\documentclass[aps,pre,showpacs,twocolumn,superscriptaddress]{revtex4-1}
\pdfoutput=1
\usepackage{amsmath,amsthm,amssymb}
\usepackage{array}
\usepackage{graphicx}
\usepackage{fancyhdr}
\usepackage{color}
\usepackage{extarrows}
\usepackage{enumerate}
\usepackage{epstopdf}
\usepackage[colorlinks,
            linkcolor=black,
            citecolor=black,
            urlcolor=blue
            ]{hyperref}
\usepackage{natbib}
\begin{document}

\title{Summing over trajectories of stochastic dynamics with multiplicative noise}

\author{Ying Tang}
\email{Corresponding author. Email: jamestang23@gmail.com}
\affiliation{Department of Physics and Astronomy, Shanghai Jiao Tong University, Shanghai 200240, China}
\affiliation{Key Laboratory of Systems Biomedicine Ministry of Education, Shanghai Center for Systems Biomedicine, Shanghai Jiao Tong University, Shanghai 200240,
China}
\author{Ruoshi Yuan}
\affiliation{School of Biomedical Engineering, Shanghai Jiao Tong University, Shanghai 200240, China}
\author{Ping Ao}
\email{Corresponding author. Email: aoping@sjtu.edu.cn}
\affiliation{Key Laboratory of Systems Biomedicine Ministry of Education, Shanghai Center for Systems Biomedicine, Shanghai Jiao Tong University, Shanghai 200240,
China}
\affiliation{Department of Physics and Astronomy, Shanghai Jiao Tong University, Shanghai 200240, China}

\date{\today}

\begin{abstract}
We demonstrate that the conventional path integral formulations generate inconsistent results exemplified by the geometric Brownian motion under the general stochastic interpretation. We thus develop a novel path integral formulation for the overdamped Langevin equation with the multiplicative noise. The present path integral leads to the corresponding Fokker-Planck equation, and naturally gives a normalized transition probability consistently in examples for general stochastic interpretations. Our result can be applied to study the fluctuation theorems and numerical calculations based on the path integral framework.

\end{abstract}
\pacs{05.40.-a, 05.10.Gg, 05.40.Jc, 05.70.Ln}
\maketitle

\section{Introduction}
The Onsager-Machlup function \cite{machlup1953fluctuations} is essential to applications such as sampling the rare events and determining the most probable path of a diffusion process \cite{Zuckerman2000Efficient,Faccioli2006Dominant,PhysRevLett.96.168101,PhysRevLett.109.150601}. To describe time-reversible dynamics, the effective action based on the symmetrical (Stratonovich's) interpretation \cite{graham1977path,arnold2000symmetric,chaichian2001path,PhysRevE.82.051104} is applied to the system with additive noise \cite{Faccioli2006Dominant,Zuckerman2000Efficient}. 
In presence of multiplicative noise, the choice of the stochastic interpretation depends on the system under study. For example, a recent experiment on a Brownian particle near a wall \cite{volpe2010influence} suggests that the system favors the anti-Ito's interpretation \cite{hanggi1978derivations,kwon2005structure,ao2008emerging,yuan2012beyond} rather than the Stratonovich's, to ensure the Boltzmann-Gibbs distribution for the final steady state. Therefore, path integral for the general stochastic interpretation is required to see which interpretation agrees with the dynamical behavior of the system under consideration.

Previous attempts on constructing path integral for the general stochastic interpretation \cite{hunt1981path,lau2007state,arenas2010functional} (called the $\alpha$-interpretation) are still controversial. The uniqueness of the action function is claimed \cite{hunt1981path}, while the action in \cite{lau2007state,arenas2010functional} depends on the stochastic interpretation. The uniqueness of the action function contradicts to the Fokker-Planck formulation that different interpretations lead to corresponding different processes \cite{gardiner2004handbook}. On the other hand, when applying the $\alpha$-interpretation path integral formula in \cite{lau2007state,arenas2010functional} to the geometric Brownian motion, we notice that the transition probability violates the conservation law.

In this paper, we provide an alternative way to construct the path integral formulation for the overdamped Langevin equation under the $\alpha$-interpretation. Through its corresponding Langevin equation of the equivalent Stratonovich's form, we obtain the path integral formulation consistent with ordinary calculus. Our main result, Eq.~(\ref{result1}), shows that the action function is not unique but $\alpha$-dependent, and can generate the $\alpha$-interpretation Fokker-Planck equation \cite{shi2012relation}. It leads to transition probabilities obeying the conservation law for general stochastic interpretations exemplified by the Ornstein-Uhlenbeck process and the geometric Brownian motion.

Our derivation on the path integral formulation demonstrates that the form of the action function and the stochastic interpretation should be consistent. Thus, for the system with additive noise, the classical Onsager-Machlup function \cite{machlup1953fluctuations} with the Ito's integration \cite{gardiner2004handbook} and the effective action \cite{graham1977path,arnold2000symmetric,chaichian2001path,PhysRevE.82.051104} with the Stratonovich's calculus are equivalent \cite{adib2008stochastic}. For the system with multiplicative noise, our action function generalizes the Onsager-Machlup function to be $\alpha$-dependent. Furthermore, as the present path integral is consistent with ordinary calculus for general stochastic interpretations, it can be applied to see whether the fluctuation theorem \cite{evans1993probability,jarzynski1997nonequilibrium,crooks1999entropy,Kim2001Fluctuation,ge2008generalized,seifert2012stochastic,PhysRevLett.111.130601} depends on the stochastic interpretation by considering the ratio of the transition probabilities of the forward and the reverse processes.

This paper is organized as follows. In Sec.~\ref{section 1}, we provide the path integral formulation and discuss its relation with the previous path integral frameworks. In Sec.~\ref{section 2}, we generate the corresponding Fokker-Planck equation from the present path integral formulation. In Sec.~\ref{section 3}, we obtain the transition probabilities for the Ornstein-Uhlenbeck process and the geometric Brownian motion under the general stochastic interpretation. In Sec.~\ref{section 4}, we summarize our work. In appendix.~\ref{appendix 1}, we list the conventional path integral frameworks and apply them to the geometric Brownian motion to show the difference with our result. In appendix.~\ref{appendix 2}, we develop an equivalent form of the path integral formulation in the main text.

\section{Path Integral Formulation}
\label{section 1}
For convenience, we start from the one dimensional overdamped Langevin equation with a multiplicative noise:
\begin{align}
\label{Langevin1}
\dot{x}=f(x)+g(x)\xi(t),
\end{align}
where $x$ denotes the position, $\dot{x}$ denotes its time derivative, $f(x)$ is the drift term and $g(x)\xi(t)$ models the stochastic force. Here, $\xi(t)$ is a Gaussian white noise with  $\langle\xi(t)\rangle=0$, $\langle\xi(t)\xi(s)\rangle=\epsilon\delta(t-s)$ and the average is taken with respect to the noise distribution. The positive constant $\epsilon$ describes the strength of the noise, corresponding to $k_{B}T$ in physical systems. For this Langevin equation, an ambiguity in choosing the integration method leads to different stochastic interpretations and a general notation is the $\alpha$-interpretation \cite{shi2012relation}. The values $\alpha=0$, $\alpha=1/2$ and $\alpha=1$ correspond to Ito's, Stratonovich's and anti-Ito's respectively.

For the Langevin equation under the $\alpha$-interpretation, by modifying the drift term, we have its equivalent Langevin equation under the Stratonovich's interpretation \cite{gardiner2004handbook}:
\begin{align}
\label{Langevin2}
\dot{x}=f(x)+\Big(\alpha-\frac{1}{2}\Big)g^{'}(x)g(x)+g(x)\xi(t),
\end{align}
where the superscript prime denotes the derivative to $x$. The advantage of using this Stratonovich's form is that ordinary calculus rule can be simply applied \cite{arenas2012hidden}. Then, this equation can be transformed to be a Langevin equation with a additive noise by a change of variable $q=H(x)$ with $H^{'}(x)=1/g(x)$ \cite{hunt1981path}:
\begin{align}
\label{Langevin4}
\dot{q}-h(q)=\xi(t),
\end{align}
where we have introduced an auxiliary function:
\begin{align}
h(q)=\frac{f\big(H^{-1}(q)\big)}{g\big(H^{-1}(q)\big)}+\Big(\alpha-\frac{1}{2}\Big)g^{'}\big(H^{-1}(q)\big).
\end{align}

To get the transition probability for Eq.~(\ref{Langevin4}), we first discretize the time into $N$ segments: $t_{0}<t_{1}<\dots<t_{N-1}<t_{N}$ with $\tau= t_{n}-t_{n-1}$ small and let $q_{n}=q(t_{n})$. For the sake of consistency, as we have chosen the equivalent Stratonovich's form, the corresponding discretized Langevin equation needs the mid-point discretization:
\begin{align}
\label{Langevin_discret}
q_{n}-q_{n-1}-\Big[\frac{h(q_{n})+h(q_{n-1})}{2}\Big]\tau=W_{n}-W_{n-1},
\end{align}
where $W(t)$ is the Wiener process given by $dW(t)=\xi(t)dt$. Thus, the Jacobian for the variable transformation between $q(t)$ and $W(t)$ is:
\begin{align}
&J\approx\exp\Big[-\frac{\tau}{2}\sum_{n=1}^{N-1}\frac{dh(q_{n})}{dq_{n}}\Big].
\end{align}

Then, with the property of the Wiener process and the Chapman-Kolmogorov equation \cite{gardiner2004handbook}, the path integral formulation for Eq.~(\ref{Langevin4}) is obtained:
\begin{align}
\label{transitional probability1}
&P(q_{N} t_{N}| q_{0} t_{0})
\notag\\&=\int^{q_{N}}_{q_{0}}\mathcal{D}q\exp\Big\{-\int^{t_{N}}_{t_{0}}\Big[\frac{1}{2\epsilon}(\dot{q}-h)^{2}+\frac{1}{2}\frac{dh}{dq}\Big]dt\Big\},
\end{align}
where $\int^{q_{N}}_{q_{0}}\mathcal{D}q\doteq\lim_{N\rightarrow\infty}\frac{1}{\sqrt{2\pi\tau\epsilon}}\prod^{N-1}_{n=1}\int\frac{dq_{n}}{\sqrt{2\pi\tau\epsilon}}$. The integral of the action function on the exponent obeys ordinary calculus due to the mid-point discretization and the last term comes from the Jacobian.

By changing the variable reversely: $x=H^{-1}(q)$ with $dx/dq=g(x)$, we get the path integral for Eq.~(\ref{Langevin1}) under the $\alpha$-interpretation:
\begin{widetext}
\begin{align}
\label{result1}
&P(x_{N} t_{N}| x_{0} t_{0})=\int^{x_{N}}_{x_{0}}\mathcal{D}x\exp\Big\{-\int^{t_{N}}_{t_{0}}\Big[\frac{1}{2g^{2}\epsilon}\Big(\dot{x}-f
-\Big(\alpha-\frac{1}{2}\Big)g^{'}g\Big)^{2}+\frac{g}{2}\Big(\frac{f}{g}+\Big(\alpha-\frac{1}{2}\Big)g^{'}\Big)^{'}\Big]dt\Big\},
\end{align}
\end{widetext}
where $\int^{x_{N}}_{x_{0}}\mathcal{D}x\doteq\lim_{N\rightarrow\infty}\frac{1}{\sqrt{2\pi\tau\epsilon}g(x_{N})}\prod^{N-1}_{n=1}\int\frac{dx_{n}}{\sqrt{2\pi\tau\epsilon}g(x_{n})}$. The action function is dimensionless because $\epsilon$ has the same dimension as energy. Though the Jacobian term comes from the measure transformation and does not belong to the conventional action part, it is usually included in the action function for applications.

The path integral formulation for the case with multiplicative noise may not be absolute continuous \cite{durr1978onsager}. However, we will show that the semi-classical method can still be applied with the present path integral to get a correct transition probability in examples, including the case with multiplicative noise. For the case with additive noise, Eq.~(\ref{result1}) degenerates to be:
\begin{align}
\label{additive1}
&P(x_{N} t_{N}| x_{0} t_{0})
\notag\\&=\int^{x_{N}}_{x_{0}}\mathcal{D}x\exp\Big\{-\int^{t_{N}}_{t_{0}}\Big[\frac{1}{2g^{2}\epsilon}(\dot{x}-f)^{2}+\frac{1}{2}f^{'}\Big]dt\Big\}.
\end{align}

We remark that though we use the equivalent Stratonovich's form above, the path integral here is for the Langevin equation under the $\alpha$-interpretation. We will also use other types of equivalent Langevin equations with the corresponding stochastic interpretation to construct the path integral formulation in appendix.~\ref{appendix 2}. The chosen stochastic interpretation in turn assigns a corresponding discretized scheme for the path integral. For example, the mid-point discretization should be used for the action function under the Stratonovich's form, and the $\alpha$-type discretization (at the point $\alpha x_{n}+(1-\alpha) x_{n-1}$ in each interval) is needed for the $\alpha$-form. These forms with the consistent stochastic interpretation are equivalent. Thus, one can choose any specific form in applications for the convenience of numerical calculations.

When doing integration for the action function, we should keep using the stochastic calculus rule consistent with its discretized scheme. If we apply the path integral here, we need to use the Stratonovich's calculus. When the path integral in \cite{lau2007state,arenas2010functional} is applied, the $\alpha$-type integration rule \cite{arenas2012hidden} is required. Taking the Ornstein-Uhlenbeck process as an example, we will show in appendix.~\ref{appendix 2} that both their path integral with the $\alpha$-type integration and Eq.~(\ref{additive1}) with ordinary calculus can automatically give a normalized transition probability. Even so, when applying their path integral to the geometric Brownian motion, the transition probabilities obtained by both ordinary calculus and the $\alpha$-type integration violate the conservation law.

The consistency of the path integral and the stochastic calculus has been noted in \cite{hunt1981path}. However, only the Langevin equation under the Ito's interpretation is considered at the beginning. Thus, their result is only for the Ito's interpretation and not for the $\alpha$-interpretation, corresponding to $\alpha=0$ in Eq.~(\ref{result1}). This explains why their result shows the uniqueness of the path integral.

\section{The Fokker-Planck Equation}
\label{section 2}
In this section, we first derive the Fokker-Planck equation from the path integral for the system with additive noise. Then, by the variable transformation, we obtain the Fokker-Planck equation of the $\alpha$-interpretation for Eq.~(\ref{Langevin1}). According to Eq.~(\ref{transitional probability1}), the transition probability in each interval is:

\begin{align}
\label{transitional probability2.2}
&P(q_{n} t_{n}| q_{n-1} t_{n-1})=\frac{1}{\sqrt{2\pi\tau\epsilon}}\exp\Big\{
\notag\\&-\frac{\tau}{2\epsilon}\Big(\frac{\Delta q_{n}}{\tau}-\frac{h(q_{n})+h(q_{n-1})}{2}\Big)^{2}-\frac{\tau}{2}\frac{dh}{dq}(q_{n-1})\Big\}
\end{align}

where $\Delta q_{n}=q_{n}-q_{n-1}$. Thus, the normalization condition is satisfied: $\int P(q_{n} t_{n}| q_{n-1} t_{n-1})dq_{n}=1$. Then, by the moment generating function \cite{wehner1987numerical}:
\begin{align}
A_{k}(q_{n-1})=\frac{(-1)^{k}}{\tau k!}\int\,dq_{n}(q_{n}-q_{n-1})^{k}P(q_{n} t_{n}| q_{n-1} t_{n-1}),
\end{align}
we calculate out the first two of $A_{k}(q_{n-1})$ as $A_{k}(q_{n-1})\approx O(\tau)$ for $k>2$:
\begin{align}
A_{1}(q_{n-1})&=-h(q_{n-1})+O(\tau),
\\A_{2}(q_{n-1})&=\frac{\epsilon}{2}+O(\tau).
\end{align}
By taking the limit $\tau\rightarrow 0$, we get the Fokker-Planck equation for Eq.~(\ref{Langevin4}):
\begin{align}
\label{Fokker-Planck2.2}
\partial_{t}\rho(q,t)&=-\partial_{q}[h(q)\rho(q,t)]+\frac{\epsilon}{2}\partial^{2}_{q}[\rho(q,t)].
\end{align}

In order to derive the Fokker-Planck equation for Eq.~(\ref{Langevin1}), we change the variable inversely: $x=H^{-1}(q)$ with $dx/dq=g(x)$. With the aid of the corresponding transformation for the moments of the Fokker-Planck equation \cite{risken1996fokker}, we have:
\begin{align}
\partial_{t}\rho(x,t)&=-\partial_{x}\Big[\Big(f(x)+\alpha g^{'}(x)g(x)\Big)\rho(x,t)\Big]
\notag\\&\quad+\frac{\epsilon}{2}\partial_x\Big[g^{2}(x)\rho(x,t)\Big],
\end{align}
which is the same as the conventional $\alpha$-interpretation Fokker-Planck equation \cite{shi2012relation}.

We emphasize that when taking the partial derivative to $x$ in the Fokker-Planck equation, we can always use ordinary calculus regardless of the interpretation adopted for the Langevin equation. In the Langevin dynamics, the expansion for a smooth function of $x$ is in orders of different time scales: $dW$, $dt$, $dW^{2}$ $dWdt$, etc. Then, $dW^{2}$ should be counted up to the order of $dt$ \cite{gardiner2004handbook}, which leads to different stochastic interpretations. However, on the level of the Fokker-Planck equation, the expansion is in orders of the space coordinates: $dx$, $dx^{2}$, etc. Thus, there is no ambiguity for the stochastic interpretation for a given Fokker-Planck equation.

\section{Examples}
\label{section 3}
\subsection{The Ornstein-Uhlenbeck process}
\label{section 3.1}
This process can be described by the Langevin equation \cite{gardiner2004handbook}:
\begin{align}
\label{Ornstein-Uhlenbeck_SDE}
\dot{x}=-kx+\sqrt{D}\xi(t),
\end{align}
where $k$, $D$ are positive constants. The temperature has been set to be a unit in examples for convenience.

We apply Eq.~(\ref{result1}):
\begin{align}
&P(x_{N} t_{N}| x_{0} t_{0})
\notag\\&=\int^{x_{N}}_{x_{0}}\mathcal{D}x\exp\Big\{-\frac{1}{2D}\int_{t_{0}}^{t_{N}}( \dot{x}+kx)^{2}dt+\frac{k\Delta t}{2}\Big\},
\end{align}
where $\Delta t=t_{N}-t_{0}$. We then get the transition probability by the semi-classical method \cite{feynman1965quantum} using ordinary calculus rule in calculation of the action function:
\begin{align}
\label{result-Ornstein}
&P(x_{N} t_{N}| x_{0} t_{0})\notag\\&=
\sqrt{\frac{k}{D\pi(1-e^{-2k\Delta{t}})}}\exp\Big[-\frac{k(x_{N}-e^{-k\Delta{t}}x_{0})^{2}}{D(1-e^{-2k\Delta{t}})}\Big].
\end{align}
If we apply the path integral in \cite{lau2007state,arenas2010functional} with the consistent $\alpha$-type integration, we obtain the same transition probability through a similar procedure. We also note that their path integral with ordinary calculus can not lead to this consistent result, Eq.~(\ref{result-Ornstein}).

\subsection{The geometric Brownian motion}
\label{section 3.2}
This process is popular in mathematical finance and recently attracts more interest in physical society \cite{PhysRevLett.110.100603}. It can be given by the following Langevin equation \cite{osendal1992stochastic}:
\begin{align}
\dot{x}= k x+\sigma x\xi(t),
\end{align}
where $ k$, $\sigma$ are positive constants. We apply Eq.~(\ref{result1}):
\begin{align}
&P(x_{N} t_{N}| x_{0} t_{0})
\notag\\&=\int^{x_{N}}_{x_{0}}\mathcal{D}x\exp\Big\{-\int_{t_{0}}^{t_{N}}\frac{[\dot{x}-kx-(\alpha-1/2)\sigma^{2}x]^{2}}{2\sigma^{2}x^{2}}dt\Big\}.
\end{align}

In the action function, to calculate the path-dependent term:
\begin{align}
\textbf{S}_{0}&=\frac{1}{2\sigma^{2}}\int_{t_{0}}^{t_{N}}\frac{\dot{x}^{2}}{x^{2}}\,dt,
\end{align}
we make a variable transformation $y=\ln x$ and then $\dot{y}=\dot{x}/x$ by ordinary calculus. Thus, through the semi-classical method \cite{feynman1965quantum}, we finally reach the result:
\begin{align}
\label{result._GBM}
&P(x_{N} t_{N}| x_{0} t_{0})=\frac{1}{\sqrt{2\pi\Delta t}\sigma x_{N}}
\notag\\&\times\exp\Big\{-\frac{\big[\ln(x_{N}/x_{0})-( k+(\alpha-1/2)\sigma^{2})\Delta t\big]^{2}}{2\sigma^{2}\Delta t}\Big\},
\end{align}
where the pre-factor $1/x_{N}$ comes from the measure transformation for $\int^{x_{N}}_{x_{0}}\mathcal{D}x$. This transition probability agrees with the result in \cite{osendal1992stochastic}, where two special cases Ito's and Stratonovich's were discussed.

When applying the path integral in \cite{lau2007state,arenas2010functional}, we derive two kinds of transition probabilities obtained by two integration rules: ordinary calculus and the $\alpha$-type integration. The detailed calculation can be found in appendix~\ref{appendix 1}. First, with ordinary calculus we have:
\begin{align}
\label{result._GBM2}
&\hat{P}(x_{N} t_{N}| x_{0} t_{0})=\frac{1}{\sqrt{2\pi\Delta t}\sigma x_{N}}
\notag\\&\times\exp\Big\{-\frac{\big[\ln(x_{N}/x_{0})-( k-\alpha\sigma^{2})\Delta t\big]^{2}}{2\sigma^{2}\Delta t}-\alpha k\Delta t\Big\}.
\end{align}
This formula violates probability conservation due to the term $\alpha k\Delta t$. Even after normalized, i.e. when $\alpha k\Delta t$ is eliminated, it still differs from Eq.~(\ref{result._GBM}).

\begin{figure*}
\includegraphics[width=1\textwidth]{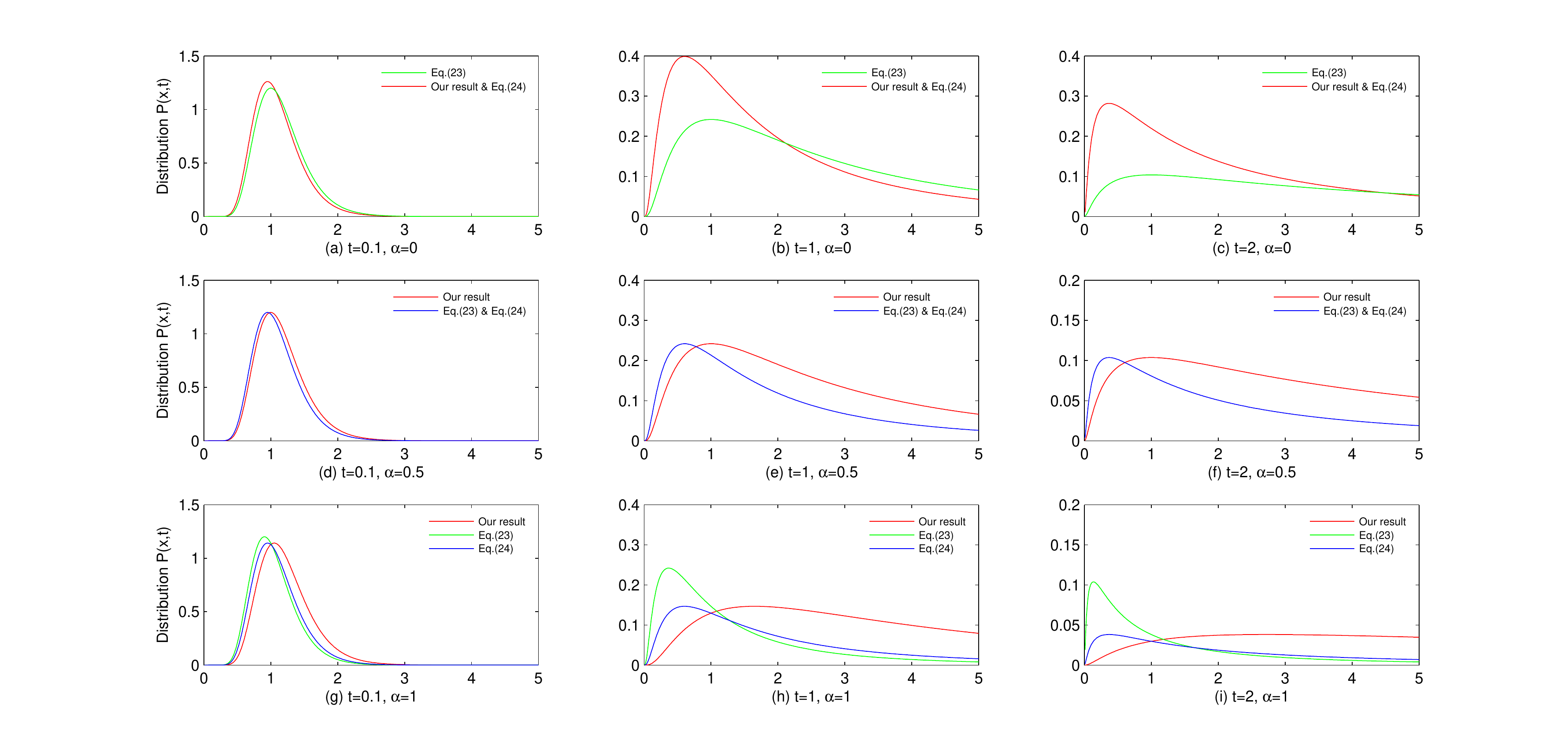}
\caption{The evolution of the probability distributions for the geometric Brownian motion (Eq.~(\ref{GBM_SDE}) with $k=1,\sigma=1$ and the delta function $\delta(x_{0}-1)$ as the initial distribution) under different stochastic interpretations: $\alpha=0$, $\alpha=1/2$, and $\alpha=1$. Different colors denote results from different path integral formulations. The three graphes in the first line correspond to Ito's interpretation ($\alpha=0$). In this case, our result Eq.~(\ref{result._GBM}) and Eq.~(\ref{result._GBM3}) coincide. The graphes in the second line correspond to Stratonovich's interpretation ($\alpha=1/2$). In this case, Eq.~(\ref{result._GBM2}) and Eq.~(\ref{result._GBM3}) coincide. In the third line, the graphes show the result under anti-Ito's interpretation ($\alpha=1$).}
\label{figure1}
\end{figure*}

Second, if we use the $\alpha$-type integration \cite{arenas2012hidden}, we have:
\begin{align}
\label{result._GBM3}
&\hat{P}(x_{N} t_{N}| x_{0} t_{0})=\frac{1}{\sqrt{2\pi\Delta t}\sigma x_{N}}
\notag\\&\times\exp\Big\{-\frac{\big[\ln(x_{N}/x_{0})-( k-\sigma^{2}/2)\Delta t\big]^{2}}{2\sigma^{2}\Delta t}-\alpha k\Delta t\Big\}.
\end{align}
Different from the Ornstein-Uhlenbeck process, this transition probability also violates probability conservation generally. After normalized, it is still different from Eq.~(\ref{result._GBM}) except under the Ito's interpretation. The comparison of transition probabilities, Eq.~(\ref{result._GBM}), Eq.~(\ref{result._GBM2}) and Eq.~(\ref{result._GBM3}) is shown in FIG.~\ref{figure1}.

\section{Conclusion}
\label{section 4}
From the overdamped Langevin equation with multiplicative noise, we have constructed the path integral formulation for the general $\alpha$-interpretation. It is convenient for applications as ordinary calculus can be applied. The corresponding $\alpha$-interpretation Fokker-Planck equation has been generated, and thus the three widely used descriptions in stochastic process are connected. Our result demonstrates the equivalence of the effective action in \cite{Faccioli2006Dominant} and the conventional Onsager-Machlup function with their corresponding stochastic integration. For the system with multiplicative noise, the present path integral generalizes the Onsager-Machlup function and obeys the conservation law for general stochastic interpretations.

For the high dimensional case with a non-singular diffusion matrix, our method can be used similarly to develop the path integral formulation. The fluctuation theorem based on the path integral here can be obtained. The forward and the reverse dynamical processes should be defined in accordance with the $\alpha$-interpretation Fokker-Planck equation. Whether or not the fluctuation theorem is related to the stochastic interpretation is an interesting topic to be explored. The influence of our result on the numerical side also remains to be discovered.

Stimulating discussions with Hong Qian, Alberto Imparato, David Cai, Andy Lau, Xiangjun Xing and Bo Yuan are gratefully acknowledged. We thank Jianhong Chen for the critical comments. This work is supported in part by the National 973 Project No.~2010CB529200 and by the Natural Science Foundation of China Projects No.~NSFC61073087 and No.~NSFC91029738. Ying Tang was partially supported by an Undergraduate Research Program in Zhiyuan College at Shanghai Jiao Tong University.

\section{Appendixes}
\appendix
\section{The conventional path integral}
\setcounter{section}{1}
\label{appendix 1}
In this appendix, we list two kinds of conventional path integral formulations for the Eq.~(\ref{Langevin1}) under the general stochastic interpretation. The first one is \cite{hunt1981path}:
\begin{align}
\label{result3}
&P(x_{N} t_{N}| x_{0} t_{0})=\int^{x_{N}}_{x_{0}}\mathcal{D}x\exp\Big\{
\notag\\&-\int^{t_{N}}_{t_{0}}\Big[\frac{1}{2g^{2}\epsilon}\Big(\dot{x}
-f+\frac{1}{2}g^{'}g\Big)^{2}+\frac{g}{2}\Big(\frac{f}{g}-\frac{1}{2}g^{'}\Big)^{'}\Big]dt\Big\},
\end{align}
where $\int^{x_{N}}_{x_{0}}\mathcal{D}x\doteq\lim_{N\rightarrow\infty}\frac{1}{\sqrt{2\pi\tau\epsilon }g(x_{N})}\prod^{N-1}_{n=1}\int\frac{dx_{n}}{\sqrt{2\pi\tau\epsilon }g(x_{n})}$ and the superscript prime denotes the derivative to $x$. According to their generation on the path integral, the action function obeys ordinary calculus.

Note that their path integral is independent of $\alpha$ and thus the authors claim the uniqueness of their action function. If applying their path integral with ordinary calculus \cite{hunt1981path}, we notice that the transition probability is always the same for any stochastic interpretation. However, it is known that for the geometric Brownian motion different stochastic interpretations lead to the corresponding different results \cite{osendal1992stochastic}. We further find that they just develop the path integral formula for the Langevin equation under Ito's interpretation. On the contrary, we start from the Langevin equation under the $\alpha$-interpretation and then the action function is not unique but $\alpha$-dependent.

The second conventional path integral is \cite{lau2007state,arenas2010functional}:
\begin{align}
\label{result2}
&\hat{P}(x_{N} t_{N}| x_{0} t_{0})=\int^{x_{N}}_{x_{0}}\mathcal{D}x
\exp\Big\{
\notag\\&-\int_{t_{0}}^{t_{N}}\!_{\alpha}\frac{1}{2g^{2}\epsilon}(\dot{x}
-f+\alpha g^{'}g)^{2}dt-\alpha\int^{t_{N}}_{t_{0}}f^{'}dt\Big\},
\end{align}
where $\int^{x_{N}}_{x_{0}}\mathcal{D}x\doteq\lim_{N\rightarrow\infty}\frac{1}{\sqrt{2\pi\tau\epsilon}g(x_{N})}\prod^{N-1}_{n=1}\int\frac{dx_{n}}{\sqrt{2\pi\tau\epsilon}g(x_{n})}$. The symbol $\int_{t_{0}}^{t_{N}}\!_{\alpha}$ means the integrand obeys the $\alpha$-type integration \cite{arenas2012hidden}: for a smooth function $F(x(t))$,
\begin{align}
\label{chain rule}
dF(x)&\approx F^{'}(x)dx+\frac{1-2\alpha}{2}F^{''}(x)g^{2}(x)dt.
\end{align}
The symbol of integral without subscript is the ordinary integral, i.e. $\int^{t_{N}}_{t_{0}}\!_{\frac{1}{2}}=\int^{t_{N}}_{t_{0}}$. Thus, we omit the subscript $1/2$ of $\int^{t_{N}}_{t_{0}}\!_{\frac{1}{2}}$ in this paper. The integral for the Jacobian term does not specify the stochastic interpretation and always obeys ordinary calculus.

We notice that it is necessary to use the $\alpha$-type integration for the action function, as their corresponding discretization is at the point $\alpha x_{n}+(1-\alpha)x_{n-1}$ in each interval. For the case with the additive noise, Eq.~(\ref{result2}) becomes:
\begin{align}
\label{additive2}
\hat{P}(x_{N} t_{N}| x_{0} t_{0})&=\int^{x_{N}}_{x_{0}}\mathcal{D}x\exp\Big\{-\int_{t_{0}}^{t_{N}}\!_{\alpha}\frac{1}{2g^{2}\epsilon}(\dot{x}-f)^{2}dt
\notag\\&\quad-\alpha\int_{t_{0}}^{t_{N}}f^{'}dt\Big\}.
\end{align}
We find that for the Ornstein-Uhlenbeck process the transition probability calculated by Eq.~(\ref{additive2}) directly with ordinary calculus is not normalized automatically. On the contrary, the corresponding $\alpha$-type integration naturally leads to a normalized transition probability. Therefore, the consistent stochastic integration is necessary to calculate the action function.

However, for the geometric Brwonian motion, we will show in the following that the transition probabilities by applying Eq.~(\ref{result2}) with both ordinary calculus and the $\alpha$-type integration are not consistent with the known result \cite{osendal1992stochastic}. Before that, we first provide the transition probability for the geometric Brownian motion under the general stochastic interpretation by generalizing the derivation in \cite{osendal1992stochastic}.

For the one dimensional geometric Brownian motion under the $\alpha$-interpretation, the equivalent equation under the Ito's interpretation is:
\begin{align}
\label{GBM_SDE}
\dot{x}=( k+\alpha\sigma^{2})x+\sigma x\xi(t).
\end{align}
The Ito's formula \cite{gardiner2004handbook}, i.e. $\alpha=0$ in Eq.~(\ref{chain rule}), tells that for $F(x)=\ln x$:
\begin{align}
dF(x)&\approx\Big( k+\alpha\sigma^{2}-\frac{1}{2}\sigma^{2}\Big)dt+\sigma dW(t).
\end{align}
As a result, the solution to Eq.~(\ref{GBM_SDE}) given initial condition $x_{0}$  at time $t_{0}$ is:
\begin{align}
x(t)=x_{0}\exp\Big\{\Big[ k+\Big(\alpha-\frac{1}{2}\Big)\sigma^{2}\Big]t+\sigma W(t)\Big\}.
\end{align}

By the distribution function of $x(t)$:
\begin{align}
&F_{X}(x_{N}t_{N}|x_{0}t_{0})
\notag\\&=P\bigg(\frac{W(\Delta t)}{\sqrt{\Delta t}}\leq\frac{[\ln(x_{N}/x_{0})-( k+(\alpha-1/2)\sigma^{2})\Delta t]}{\sigma\sqrt{\Delta t}}\bigg),
\end{align}
we get the transition probability:
\begin{align}
\label{result_GMB1}
&P(x_{N}t_{N}|x_{0}t_{0})=\frac{1}{\sqrt{2\pi\Delta t}\sigma x_{N}}
\notag\\&\times\exp\Big\{-\frac{\big[\ln(x_{N}/x_{0})-( k+(\alpha-1/2)\sigma^{2})\Delta t\big]^{2}}{2\sigma^{2}\Delta t}\Big\}.
\end{align}
It is consistent with the previous result in \cite{osendal1992stochastic}, where Ito's and Stratonovich's cases were discussed.

Now, we apply Eq.~(\ref{result2}) and have the action function:
\begin{align}
\hat{\textbf{S}}&=\frac{1}{2\sigma^{2}}\int_{t_{0}}^{t_{N}}\!_{\alpha}\Big\{\frac{\dot{x}^{2}}{x^{2}}-2( k-\alpha\sigma^{2})\frac{\dot{x}}{x}+( k-\alpha\sigma^{2})^{2}\Big\}dt
\notag\\&\quad+\alpha k\Delta t.
\end{align}
We then use two different integration rules separately. First, we use ordinary calculus. For the first term of the action function, we make a variable transformation $y=\ln x$ and thus $\dot{y}=\dot{x}/x$. Then, with the semi-classical method on a free particle \cite{feynman1965quantum}, we have Eq.~(\ref{result._GBM2}).

Second, if we use the $\alpha$-type integration for the action function, then
\begin{align}
\frac{1}{2\sigma^{2}}\int_{t_{0}}^{t_{N}}\!_{\alpha}\Big(\frac{\dot{x}}{x}\Big)dt&=\frac{1}{2\sigma^{2}}\Big[\ln\Big(\frac{x_{N}}{x_{0}}\Big)+\Big(\frac{1}{2}-\alpha\Big)\sigma^{2}\Delta t\Big].
\end{align}
Besides, for the variable transformation $y=\ln x$, we should have $\dot{y}=\dot{x}/x-(1/2-\alpha)\sigma^{2}$ and finally obtain Eq.~(\ref{result._GBM3}) after similar procedure.

\section{Path integral formulation of the equivalent $\alpha$-form}
\label{appendix 2}
In this appendix, we provide another way to develop the path integral formulation starting from the one dimensional Langevin equation under the $\alpha$-interpretation. Instead of modifying the drift term and writing down its equivalent Langevin equation in Stratonovich's form, we directly use Eq.~(\ref{Langevin1}). The $\alpha$-type chain rule, Eq.~(\ref{chain rule}), should be applied to do the variable transformation, $q=H(x)$:
\begin{align}
dH(x)&\approx H^{'}(x)dx+\frac{1-2\alpha}{2}H^{''}(x)g^{2}(x)dt,
\end{align}
where the superscript prime denotes the derivative to $x$. Then, with $H^{'}(x)=1/g(x)$ and $H^{''}(x)=-g^{'}(x)/g^{2}(x)$ Eq.~(\ref{Langevin1}) can be transformed to be a Langevin equation with the additive noise:
\begin{align}
\label{LangevinB.4}
\dot{q}-h(q)=\xi(t),
\end{align}
with
\begin{align}
h(q)=\frac{f\big(H^{-1}(q)\big)}{g\big(H^{-1}(q)\big)}+\Big(\alpha-\frac{1}{2}\Big)g^{'}\big(H^{-1}(q)\big).
\end{align}

For the sake of consistency, as we have chosen the $\alpha$-interpretation, the corresponding discretized Langevin equation should be:
\begin{align}
\label{Langevin_discret B}
q_{n}-q_{n-1}-[\alpha h(q_{n})+(1-\alpha)h(q_{n-1})]\tau=W_{n}-W_{n-1}.
\end{align}
Thus, the Jacobian for the variable transformation between $q(t)$ and $W(t)$ becomes:
\begin{align}
J\approx\exp\Big[-\alpha\tau\sum_{n=1}^{N-1}\frac{dh(q_{n})}{dq_{n}}\Big].
\end{align}

Then, with the property of Wiener process and the Chapman-Kolmogorov equation \cite{hunt1981path}, the transition probability is obtained:
\begin{align}
\label{transitional probabilityB.1}
P(q_{N} t_{N}| q_{0} t_{0})&=\int^{q_{N}}_{q_{0}}\mathcal{D}q\exp\Big\{-\int^{t_{N}}_{t_{0}}\!_{\alpha}\frac{1}{2\epsilon}[\dot{q}-h(q)]^{2}dt
\notag\\&\quad-\alpha\int^{t_{N}}_{t_{0}}\frac{dh(q)}{dq}dt\Big\}.
\end{align}
To get the transition probability $P(x_{N} t_{N}| x_{0} t_{0})$ for Eq.~(\ref{Langevin1}), we change the variable reversely:
\begin{align}
\label{resultB.1}
P(x_{N} t_{N}| x_{0} t_{0})&=\int^{x_{N}}_{x_{0}}\mathcal{D}x\exp\Big\{
-\int^{t_{N}}_{t_{0}}\!_{\alpha}\frac{1}{2g^{2}\epsilon}(\dot{x}-f)^{2}dt
\notag\\&\quad-\alpha\int^{t_{N}}_{t_{0}}g\Big[\frac{f}{g}+\Big(\alpha-\frac{1}{2}\Big)g^{'}\Big]^{'}dt\Big\}.
\end{align}
The different forms of the present path integral come from the equivalent Langevin equation under different stochastic interpretations. With the consistent calculus rule, all the forms are equivalent. The corresponding integration rule should be applied with the specific form of the present path integral. Therefore, one can conveniently choose the path integral form with the integration rule for the problem considered.

For the additive noise cases, Eq.~(\ref{resultB.1}) becomes
\begin{align}
\label{additive3}
P(x_{N} t_{N}| x_{0} t_{0})&=\int^{x_{N}}_{x_{0}}\mathcal{D}x\exp\Big\{-\int_{t_{0}}^{t_{N}}\!_{\alpha}\frac{1}{2g^{2}\epsilon}(\dot{x}-f)^{2}dt
\notag\\&\quad-\alpha\int_{t_{0}}^{t_{N}}f^{'}dt\Big\},
\end{align}
which is the same as Eq.~(\ref{additive2}). This demonstrates that the present path integral formulation and that in \cite{lau2007state,arenas2010functional} show no difference for the additive noise cases. It should be emphasized that though Eq.~(\ref{additive3}) explicitly contains an $\alpha$-term, it is independent with $\alpha$ after integration. Because the integral $\int_{t_{0}}^{t_{N}}\!_{\alpha}$ should be done through the $\alpha$-type integration, which will eliminate the $\alpha$-term on the exponent.

\bibliography{bib}

\begin{thebibliography}{34}%
\makeatletter
\providecommand \@ifxundefined [1]{%
 \@ifx{#1\undefined}
}%
\providecommand \@ifnum [1]{%
 \ifnum #1\expandafter \@firstoftwo
 \else \expandafter \@secondoftwo
 \fi
}%
\providecommand \@ifx [1]{%
 \ifx #1\expandafter \@firstoftwo
 \else \expandafter \@secondoftwo
 \fi
}%
\providecommand \natexlab [1]{#1}%
\providecommand \enquote  [1]{``#1''}%
\providecommand \bibnamefont  [1]{#1}%
\providecommand \bibfnamefont [1]{#1}%
\providecommand \citenamefont [1]{#1}%
\providecommand \href@noop [0]{\@secondoftwo}%
\providecommand \href [0]{\begingroup \@sanitize@url \@href}%
\providecommand \@href[1]{\@@startlink{#1}\@@href}%
\providecommand \@@href[1]{\endgroup#1\@@endlink}%
\providecommand \@sanitize@url [0]{\catcode `\\12\catcode `\$12\catcode
  `\&12\catcode `\#12\catcode `\^12\catcode `\_12\catcode `\%12\relax}%
\providecommand \@@startlink[1]{}%
\providecommand \@@endlink[0]{}%
\providecommand \url  [0]{\begingroup\@sanitize@url \@url }%
\providecommand \@url [1]{\endgroup\@href {#1}{\urlprefix }}%
\providecommand \urlprefix  [0]{URL }%
\providecommand \Eprint [0]{\href }%
\providecommand \doibase [0]{http://dx.doi.org/}%
\providecommand \selectlanguage [0]{\@gobble}%
\providecommand \bibinfo  [0]{\@secondoftwo}%
\providecommand \bibfield  [0]{\@secondoftwo}%
\providecommand \translation [1]{[#1]}%
\providecommand \BibitemOpen [0]{}%
\providecommand \bibitemStop [0]{}%
\providecommand \bibitemNoStop [0]{.\EOS\space}%
\providecommand \EOS [0]{\spacefactor3000\relax}%
\providecommand \BibitemShut  [1]{\csname bibitem#1\endcsname}%
\let\auto@bib@innerbib\@empty
\bibitem [{\citenamefont {Machlup}\ and\ \citenamefont
  {Onsager}(1953)}]{machlup1953fluctuations}%
  \BibitemOpen
  \bibfield  {author} {\bibinfo {author} {\bibfnamefont {S.}~\bibnamefont
  {Machlup}}\ and\ \bibinfo {author} {\bibfnamefont {L.}~\bibnamefont
  {Onsager}},\ }\href {\doibase 10.1103/PhysRev.91.1505} {\bibfield  {journal}
  {\bibinfo  {journal} {Phys. Rev.}\ }\textbf {\bibinfo {volume} {91}},\
  \bibinfo {pages} {1512} (\bibinfo {year} {1953})}\BibitemShut {NoStop}%
\bibitem [{\citenamefont {Zuckerman}\ and\ \citenamefont
  {Woolf}(2000)}]{Zuckerman2000Efficient}%
  \BibitemOpen
  \bibfield  {author} {\bibinfo {author} {\bibfnamefont {D.~M.}\ \bibnamefont
  {Zuckerman}}\ and\ \bibinfo {author} {\bibfnamefont {T.~B.}\ \bibnamefont
  {Woolf}},\ }\href {\doibase 10.1103/PhysRevE.63.016702} {\bibfield  {journal}
  {\bibinfo  {journal} {Phys. Rev. E}\ }\textbf {\bibinfo {volume} {63}},\
  \bibinfo {pages} {016702} (\bibinfo {year} {2000})}\BibitemShut {NoStop}%
\bibitem [{\citenamefont {Faccioli}\ \emph {et~al.}(2006)\citenamefont
  {Faccioli}, \citenamefont {Sega}, \citenamefont {Pederiva},\ and\
  \citenamefont {Orland}}]{Faccioli2006Dominant}%
  \BibitemOpen
  \bibfield  {author} {\bibinfo {author} {\bibfnamefont {P.}~\bibnamefont
  {Faccioli}}, \bibinfo {author} {\bibfnamefont {M.}~\bibnamefont {Sega}},
  \bibinfo {author} {\bibfnamefont {F.}~\bibnamefont {Pederiva}}, \ and\
  \bibinfo {author} {\bibfnamefont {H.}~\bibnamefont {Orland}},\ }\href
  {\doibase 10.1103/PhysRevLett.97.108101} {\bibfield  {journal} {\bibinfo
  {journal} {Phys. Rev. Lett.}\ }\textbf {\bibinfo {volume} {97}},\ \bibinfo
  {pages} {108101} (\bibinfo {year} {2006})}\BibitemShut {NoStop}%
\bibitem [{\citenamefont {Wang}\ \emph {et~al.}(2006)\citenamefont {Wang},
  \citenamefont {Zhang}, \citenamefont {Lu},\ and\ \citenamefont
  {Wang}}]{PhysRevLett.96.168101}%
  \BibitemOpen
  \bibfield  {author} {\bibinfo {author} {\bibfnamefont {J.}~\bibnamefont
  {Wang}}, \bibinfo {author} {\bibfnamefont {K.}~\bibnamefont {Zhang}},
  \bibinfo {author} {\bibfnamefont {H.}~\bibnamefont {Lu}}, \ and\ \bibinfo
  {author} {\bibfnamefont {E.}~\bibnamefont {Wang}},\ }\href {\doibase
  10.1103/PhysRevLett.96.168101} {\bibfield  {journal} {\bibinfo  {journal}
  {Phys. Rev. Lett.}\ }\textbf {\bibinfo {volume} {96}},\ \bibinfo {pages}
  {168101} (\bibinfo {year} {2006})}\BibitemShut {NoStop}%
\bibitem [{\citenamefont {Gobbo}\ \emph {et~al.}(2012)\citenamefont {Gobbo},
  \citenamefont {Laio}, \citenamefont {Maleki},\ and\ \citenamefont
  {Baroni}}]{PhysRevLett.109.150601}%
  \BibitemOpen
  \bibfield  {author} {\bibinfo {author} {\bibfnamefont {G.}~\bibnamefont
  {Gobbo}}, \bibinfo {author} {\bibfnamefont {A.}~\bibnamefont {Laio}},
  \bibinfo {author} {\bibfnamefont {A.}~\bibnamefont {Maleki}}, \ and\ \bibinfo
  {author} {\bibfnamefont {S.}~\bibnamefont {Baroni}},\ }\href {\doibase
  10.1103/PhysRevLett.109.150601} {\bibfield  {journal} {\bibinfo  {journal}
  {Phys. Rev. Lett.}\ }\textbf {\bibinfo {volume} {109}},\ \bibinfo {pages}
  {150601} (\bibinfo {year} {2012})}\BibitemShut {NoStop}%
\bibitem [{\citenamefont {Graham}(1977)}]{graham1977path}%
  \BibitemOpen
  \bibfield  {author} {\bibinfo {author} {\bibfnamefont {R.}~\bibnamefont
  {Graham}},\ }\href {http://link.springer.com/article/10.1007/BF01312935}
  {\bibfield  {journal} {\bibinfo  {journal} {Z. Physik B}\ }\textbf {\bibinfo
  {volume} {26}},\ \bibinfo {pages} {281} (\bibinfo {year} {1977})}\BibitemShut
  {NoStop}%
\bibitem [{\citenamefont {Arnold}(2000)}]{arnold2000symmetric}%
  \BibitemOpen
  \bibfield  {author} {\bibinfo {author} {\bibfnamefont {P.}~\bibnamefont
  {Arnold}},\ }\href {\doibase 10.1103/PhysRevE.61.6099} {\bibfield  {journal}
  {\bibinfo  {journal} {Phys. Rev. E}\ }\textbf {\bibinfo {volume} {61}},\
  \bibinfo {pages} {6099} (\bibinfo {year} {2000})}\BibitemShut {NoStop}%
\bibitem [{\citenamefont {Chaichian}\ and\ \citenamefont
  {Demichev}(2001)}]{chaichian2001path}%
  \BibitemOpen
  \bibfield  {author} {\bibinfo {author} {\bibfnamefont {M.}~\bibnamefont
  {Chaichian}}\ and\ \bibinfo {author} {\bibfnamefont {A.}~\bibnamefont
  {Demichev}},\ }\href@noop {} {\emph {\bibinfo {title} {Stochastic Processes
  and Quantum Mechanics, Path Integrals in Physics}}},\ Vol.~\bibinfo {volume}
  {I.}\ (\bibinfo  {publisher} {IOP, Bristol},\ \bibinfo {year}
  {2001})\BibitemShut {NoStop}%
\bibitem [{\citenamefont {Chatterjee}\ and\ \citenamefont
  {Cherayil}(2010)}]{PhysRevE.82.051104}%
  \BibitemOpen
  \bibfield  {author} {\bibinfo {author} {\bibfnamefont {D.}~\bibnamefont
  {Chatterjee}}\ and\ \bibinfo {author} {\bibfnamefont {B.~J.}\ \bibnamefont
  {Cherayil}},\ }\href {\doibase 10.1103/PhysRevE.82.051104} {\bibfield
  {journal} {\bibinfo  {journal} {Phys. Rev. E}\ }\textbf {\bibinfo {volume}
  {82}},\ \bibinfo {pages} {051104} (\bibinfo {year} {2010})}\BibitemShut
  {NoStop}%
\bibitem [{\citenamefont {Volpe}\ \emph {et~al.}(2010)\citenamefont {Volpe},
  \citenamefont {Helden}, \citenamefont {Brettschneider}, \citenamefont
  {Wehr},\ and\ \citenamefont {Bechinger}}]{volpe2010influence}%
  \BibitemOpen
  \bibfield  {author} {\bibinfo {author} {\bibfnamefont {G.}~\bibnamefont
  {Volpe}}, \bibinfo {author} {\bibfnamefont {L.}~\bibnamefont {Helden}},
  \bibinfo {author} {\bibfnamefont {T.}~\bibnamefont {Brettschneider}},
  \bibinfo {author} {\bibfnamefont {J.}~\bibnamefont {Wehr}}, \ and\ \bibinfo
  {author} {\bibfnamefont {C.}~\bibnamefont {Bechinger}},\ }\href {\doibase
  10.1103/PhysRevLett.104.170602} {\bibfield  {journal} {\bibinfo  {journal}
  {Phys. Rev. Lett.}\ }\textbf {\bibinfo {volume} {104}},\ \bibinfo {pages}
  {170602} (\bibinfo {year} {2010})}\BibitemShut {NoStop}%
\bibitem [{\citenamefont {H{\"a}nggi}(1978)}]{hanggi1978derivations}%
  \BibitemOpen
  \bibfield  {author} {\bibinfo {author} {\bibfnamefont {P.}~\bibnamefont
  {H{\"a}nggi}},\ }\href {\doibase 10.1007/BF01323672} {\bibfield  {journal}
  {\bibinfo  {journal} {Eur. Phys. J. B}\ }\textbf {\bibinfo {volume} {30}},\
  \bibinfo {pages} {85} (\bibinfo {year} {1978})}\BibitemShut {NoStop}%
\bibitem [{\citenamefont {Kwon}\ \emph {et~al.}(2005)\citenamefont {Kwon},
  \citenamefont {Ao},\ and\ \citenamefont {Thouless}}]{kwon2005structure}%
  \BibitemOpen
  \bibfield  {author} {\bibinfo {author} {\bibfnamefont {C.}~\bibnamefont
  {Kwon}}, \bibinfo {author} {\bibfnamefont {P.}~\bibnamefont {Ao}}, \ and\
  \bibinfo {author} {\bibfnamefont {D.~J.}\ \bibnamefont {Thouless}},\ }\href
  {http://www.pnas.org/content/102/37/13029.short} {\bibfield  {journal}
  {\bibinfo  {journal} {Proc. Natl. Acad. Sci. USA}\ }\textbf {\bibinfo
  {volume} {102}},\ \bibinfo {pages} {13029} (\bibinfo {year}
  {2005})}\BibitemShut {NoStop}%
\bibitem [{\citenamefont {Ao}(2008)}]{ao2008emerging}%
  \BibitemOpen
  \bibfield  {author} {\bibinfo {author} {\bibfnamefont {P.}~\bibnamefont
  {Ao}},\ }\href {\doibase 10.1088/0253-6102/49/5/01} {\bibfield  {journal}
  {\bibinfo  {journal} {Commun. Theor. Phys.}\ }\textbf {\bibinfo {volume}
  {49}},\ \bibinfo {pages} {1073} (\bibinfo {year} {2008})}\BibitemShut
  {NoStop}%
\bibitem [{\citenamefont {Yuan}\ and\ \citenamefont
  {Ao}(2012)}]{yuan2012beyond}%
  \BibitemOpen
  \bibfield  {author} {\bibinfo {author} {\bibfnamefont {R.}~\bibnamefont
  {Yuan}}\ and\ \bibinfo {author} {\bibfnamefont {P.}~\bibnamefont {Ao}},\
  }\href {http://iopscience.iop.org/1742-5468/2012/07/P07010} {\bibfield
  {journal} {\bibinfo  {journal} {J. Stat. Mech.}\ }\textbf {\bibinfo {volume}
  {2012}},\ \bibinfo {pages} {P07010} (\bibinfo {year} {2012})}\BibitemShut
  {NoStop}%
\bibitem [{\citenamefont {Hunt}\ and\ \citenamefont
  {Ross}(1981)}]{hunt1981path}%
  \BibitemOpen
  \bibfield  {author} {\bibinfo {author} {\bibfnamefont {K.~L.}\ \bibnamefont
  {Hunt}}\ and\ \bibinfo {author} {\bibfnamefont {J.}~\bibnamefont {Ross}},\
  }\href {http://dx.doi.org/10.1063/1.442098} {\bibfield  {journal} {\bibinfo
  {journal} {J. Chem. Phys.}\ }\textbf {\bibinfo {volume} {75}},\ \bibinfo
  {pages} {976} (\bibinfo {year} {1981})}\BibitemShut {NoStop}%
\bibitem [{\citenamefont {Lau}\ and\ \citenamefont
  {Lubensky}(2007)}]{lau2007state}%
  \BibitemOpen
  \bibfield  {author} {\bibinfo {author} {\bibfnamefont {A.~W.~C.}\
  \bibnamefont {Lau}}\ and\ \bibinfo {author} {\bibfnamefont {T.~C.}\
  \bibnamefont {Lubensky}},\ }\href {\doibase 10.1103/PhysRevE.76.011123}
  {\bibfield  {journal} {\bibinfo  {journal} {Phys. Rev. E}\ }\textbf {\bibinfo
  {volume} {76}},\ \bibinfo {pages} {011123} (\bibinfo {year}
  {2007})}\BibitemShut {NoStop}%
\bibitem [{\citenamefont {Arenas}\ and\ \citenamefont
  {Barci}(2010)}]{arenas2010functional}%
  \BibitemOpen
  \bibfield  {author} {\bibinfo {author} {\bibfnamefont {Z.~G.}\ \bibnamefont
  {Arenas}}\ and\ \bibinfo {author} {\bibfnamefont {D.~G.}\ \bibnamefont
  {Barci}},\ }\href {\doibase 10.1103/PhysRevE.81.051113} {\bibfield  {journal}
  {\bibinfo  {journal} {Phys. Rev. E}\ }\textbf {\bibinfo {volume} {81}},\
  \bibinfo {pages} {051113} (\bibinfo {year} {2010})}\BibitemShut {NoStop}%
\bibitem [{\citenamefont {Gardiner}(2004)}]{gardiner2004handbook}%
  \BibitemOpen
  \bibfield  {author} {\bibinfo {author} {\bibfnamefont {C.~W.}\ \bibnamefont
  {Gardiner}},\ }\href@noop {} {\emph {\bibinfo {title} {Handbook of Stochastic
  Methods}}}\ (\bibinfo  {publisher} {Springer, Berlin},\ \bibinfo {year}
  {2004})\BibitemShut {NoStop}%
\bibitem [{\citenamefont {Shi}\ \emph {et~al.}(2012)\citenamefont {Shi},
  \citenamefont {Chen}, \citenamefont {Yuan}, \citenamefont {Yuan},\ and\
  \citenamefont {Ao}}]{shi2012relation}%
  \BibitemOpen
  \bibfield  {author} {\bibinfo {author} {\bibfnamefont {J.}~\bibnamefont
  {Shi}}, \bibinfo {author} {\bibfnamefont {T.}~\bibnamefont {Chen}}, \bibinfo
  {author} {\bibfnamefont {R.}~\bibnamefont {Yuan}}, \bibinfo {author}
  {\bibfnamefont {B.}~\bibnamefont {Yuan}}, \ and\ \bibinfo {author}
  {\bibfnamefont {P.}~\bibnamefont {Ao}},\ }\href {\doibase
  10.1007/s10955-012-0532-8} {\bibfield  {journal} {\bibinfo  {journal} {J.
  Stat. Phys.}\ }\textbf {\bibinfo {volume} {148}},\ \bibinfo {pages} {579}
  (\bibinfo {year} {2012})}\BibitemShut {NoStop}%
\bibitem [{\citenamefont {Adib}(2008)}]{adib2008stochastic}%
  \BibitemOpen
  \bibfield  {author} {\bibinfo {author} {\bibfnamefont {A.~B.}\ \bibnamefont
  {Adib}},\ }\href {http://pubs.acs.org/doi/abs/10.1021/jp0751458} {\bibfield
  {journal} {\bibinfo  {journal} {J. Phys. Chem. B}\ }\textbf {\bibinfo
  {volume} {112}},\ \bibinfo {pages} {5910} (\bibinfo {year}
  {2008})}\BibitemShut {NoStop}%
\bibitem [{\citenamefont {Evans}\ \emph {et~al.}(1993)\citenamefont {Evans},
  \citenamefont {Cohen},\ and\ \citenamefont {Morriss}}]{evans1993probability}%
  \BibitemOpen
  \bibfield  {author} {\bibinfo {author} {\bibfnamefont {D.~J.}\ \bibnamefont
  {Evans}}, \bibinfo {author} {\bibfnamefont {E.~G.~D.}\ \bibnamefont {Cohen}},
  \ and\ \bibinfo {author} {\bibfnamefont {G.~P.}\ \bibnamefont {Morriss}},\
  }\href {\doibase 10.1103/PhysRevLett.71.2401} {\bibfield  {journal} {\bibinfo
   {journal} {Phys. Rev. Lett.}\ }\textbf {\bibinfo {volume} {71}},\ \bibinfo
  {pages} {2401} (\bibinfo {year} {1993})}\BibitemShut {NoStop}%
\bibitem [{\citenamefont {Jarzynski}(1997)}]{jarzynski1997nonequilibrium}%
  \BibitemOpen
  \bibfield  {author} {\bibinfo {author} {\bibfnamefont {C.}~\bibnamefont
  {Jarzynski}},\ }\href {\doibase 10.1103/PhysRevLett.78.2690} {\bibfield
  {journal} {\bibinfo  {journal} {Phys. Rev. Lett.}\ }\textbf {\bibinfo
  {volume} {78}},\ \bibinfo {pages} {2690} (\bibinfo {year}
  {1997})}\BibitemShut {NoStop}%
\bibitem [{\citenamefont {Crooks}(1999)}]{crooks1999entropy}%
  \BibitemOpen
  \bibfield  {author} {\bibinfo {author} {\bibfnamefont {G.~E.}\ \bibnamefont
  {Crooks}},\ }\href {\doibase 10.1103/PhysRevE.60.2721} {\bibfield  {journal}
  {\bibinfo  {journal} {Phys. Rev. E}\ }\textbf {\bibinfo {volume} {60}},\
  \bibinfo {pages} {2721} (\bibinfo {year} {1999})}\BibitemShut {NoStop}%
\bibitem [{\citenamefont {Kim}\ and\ \citenamefont
  {Qian}(2007)}]{Kim2001Fluctuation}%
  \BibitemOpen
  \bibfield  {author} {\bibinfo {author} {\bibfnamefont {K.~H.}\ \bibnamefont
  {Kim}}\ and\ \bibinfo {author} {\bibfnamefont {H.}~\bibnamefont {Qian}},\
  }\href {\doibase 10.1103/PhysRevE.75.022102} {\bibfield  {journal} {\bibinfo
  {journal} {Phys. Rev. E}\ }\textbf {\bibinfo {volume} {75}},\ \bibinfo
  {pages} {022102} (\bibinfo {year} {2007})}\BibitemShut {NoStop}%
\bibitem [{\citenamefont {Ge}\ and\ \citenamefont
  {Jiang}(2008)}]{ge2008generalized}%
  \BibitemOpen
  \bibfield  {author} {\bibinfo {author} {\bibfnamefont {H.}~\bibnamefont
  {Ge}}\ and\ \bibinfo {author} {\bibfnamefont {D.-Q.}\ \bibnamefont {Jiang}},\
  }\href {http://link.springer.com/article/10.1007/s10955-008-9520-4}
  {\bibfield  {journal} {\bibinfo  {journal} {J. Stat. Phys.}\ }\textbf
  {\bibinfo {volume} {131}},\ \bibinfo {pages} {675} (\bibinfo {year}
  {2008})}\BibitemShut {NoStop}%
\bibitem [{\citenamefont {Seifert}(2012)}]{seifert2012stochastic}%
  \BibitemOpen
  \bibfield  {author} {\bibinfo {author} {\bibfnamefont {U.}~\bibnamefont
  {Seifert}},\ }\href {\doibase doi:10.1088/0034-4885/75/12/126001} {\bibfield
  {journal} {\bibinfo  {journal} {Rep. Prog. Phys.}\ }\textbf {\bibinfo
  {volume} {75}},\ \bibinfo {pages} {126001} (\bibinfo {year}
  {2012})}\BibitemShut {NoStop}%
\bibitem [{\citenamefont {Noh}\ \emph {et~al.}(2013)\citenamefont {Noh},
  \citenamefont {Kwon},\ and\ \citenamefont {Park}}]{PhysRevLett.111.130601}%
  \BibitemOpen
  \bibfield  {author} {\bibinfo {author} {\bibfnamefont {J.~D.}\ \bibnamefont
  {Noh}}, \bibinfo {author} {\bibfnamefont {C.}~\bibnamefont {Kwon}}, \ and\
  \bibinfo {author} {\bibfnamefont {H.}~\bibnamefont {Park}},\ }\href {\doibase
  10.1103/PhysRevLett.111.130601} {\bibfield  {journal} {\bibinfo  {journal}
  {Phys. Rev. Lett.}\ }\textbf {\bibinfo {volume} {111}},\ \bibinfo {pages}
  {130601} (\bibinfo {year} {2013})}\BibitemShut {NoStop}%
\bibitem [{\citenamefont {Arenas}\ and\ \citenamefont
  {Barci}(2012)}]{arenas2012hidden}%
  \BibitemOpen
  \bibfield  {author} {\bibinfo {author} {\bibfnamefont {Z.~G.}\ \bibnamefont
  {Arenas}}\ and\ \bibinfo {author} {\bibfnamefont {D.~G.}\ \bibnamefont
  {Barci}},\ }\href {http://stacks.iop.org/1742-5468/2012/i=12/a=P12005}
  {\bibfield  {journal} {\bibinfo  {journal} {J. Stat. Mech.}\ }\textbf
  {\bibinfo {volume} {2012}},\ \bibinfo {pages} {P12005} (\bibinfo {year}
  {2012})}\BibitemShut {NoStop}%
\bibitem [{\citenamefont {D{\"u}rr}\ and\ \citenamefont
  {Bach}(1978)}]{durr1978onsager}%
  \BibitemOpen
  \bibfield  {author} {\bibinfo {author} {\bibfnamefont {D.}~\bibnamefont
  {D{\"u}rr}}\ and\ \bibinfo {author} {\bibfnamefont {A.}~\bibnamefont
  {Bach}},\ }\href {http://link.springer.com/article/10.1007/BF01609446}
  {\bibfield  {journal} {\bibinfo  {journal} {Comm. Math. Phys.}\ }\textbf
  {\bibinfo {volume} {60}},\ \bibinfo {pages} {153} (\bibinfo {year}
  {1978})}\BibitemShut {NoStop}%
\bibitem [{\citenamefont {Wehner}\ and\ \citenamefont
  {Wolfer}(1987)}]{wehner1987numerical}%
  \BibitemOpen
  \bibfield  {author} {\bibinfo {author} {\bibfnamefont {M.~F.}\ \bibnamefont
  {Wehner}}\ and\ \bibinfo {author} {\bibfnamefont {W.~G.}\ \bibnamefont
  {Wolfer}},\ }\href {\doibase 10.1103/PhysRevA.35.1795} {\bibfield  {journal}
  {\bibinfo  {journal} {Phys. Rev. A}\ }\textbf {\bibinfo {volume} {35}},\
  \bibinfo {pages} {1795} (\bibinfo {year} {1987})}\BibitemShut {NoStop}%
\bibitem [{\citenamefont {Risken}(1996)}]{risken1996fokker}%
  \BibitemOpen
  \bibfield  {author} {\bibinfo {author} {\bibfnamefont {H.}~\bibnamefont
  {Risken}},\ }\href@noop {} {\emph {\bibinfo {title} {The Fokker-Planck
  Equation: Methods of Solution and Applications}}},\ Vol.~\bibinfo {volume}
  {18}\ (\bibinfo  {publisher} {Springer Verlag},\ \bibinfo {year}
  {1996})\BibitemShut {NoStop}%
\bibitem [{\citenamefont {Feynman}\ and\ \citenamefont
  {Hibbs}(1965)}]{feynman1965quantum}%
  \BibitemOpen
  \bibfield  {author} {\bibinfo {author} {\bibfnamefont {R.~P.}\ \bibnamefont
  {Feynman}}\ and\ \bibinfo {author} {\bibfnamefont {A.}~\bibnamefont
  {Hibbs}},\ }\href@noop {} {\emph {\bibinfo {title} {Quantum Mechanics and
  Path Integrals}}}\ (\bibinfo  {publisher} {McGraw-Hill, New York},\ \bibinfo
  {year} {1965})\BibitemShut {NoStop}%
\bibitem [{\citenamefont {Peters}\ and\ \citenamefont
  {Klein}(2013)}]{PhysRevLett.110.100603}%
  \BibitemOpen
  \bibfield  {author} {\bibinfo {author} {\bibfnamefont {O.}~\bibnamefont
  {Peters}}\ and\ \bibinfo {author} {\bibfnamefont {W.}~\bibnamefont {Klein}},\
  }\href {\doibase 10.1103/PhysRevLett.110.100603} {\bibfield  {journal}
  {\bibinfo  {journal} {Phys. Rev. Lett.}\ }\textbf {\bibinfo {volume} {110}},\
  \bibinfo {pages} {100603} (\bibinfo {year} {2013})}\BibitemShut {NoStop}%
\bibitem [{\citenamefont {{\O}ksendal}(1992)}]{osendal1992stochastic}%
  \BibitemOpen
  \bibfield  {author} {\bibinfo {author} {\bibfnamefont {B.}~\bibnamefont
  {{\O}ksendal}},\ }\href@noop {} {\emph {\bibinfo {title} {Stochastic
  Differential Equations: an Introduction with Applications}}},\ \bibinfo
  {edition} {3rd}\ ed.\ (\bibinfo  {publisher} {Springer, Berlin},\ \bibinfo
  {year} {1992})\BibitemShut {NoStop}%
\end{thebibliography}%
\end{document}